\documentclass[8pt]{article}

\usepackage{cite}

\begin{document}

\title{Embeddings for  General Relativity}
\author{J. Ponce de Leon\thanks{E-mail:   	 jpdel1@hotmail.com}\\ Laboratory of Theoretical Physics, Department of Physics\\ 
University of Puerto Rico, P.O. Box 23343, San Juan, \\ PR 00931, USA} 
\date{September 01,  2015}

\maketitle

\begin{abstract}

We present a systematic approach to embed $n$-dimensional vacuum general relativity     
in an $(n + 1)$-dimensional pseudo-Riemannian spacetime whose source is either a (non)zero cosmological constant or  a scalar field minimally-coupled to Einstein gravity.  Our approach allows us to  generalize a number of results discussed in the literature. 
We construct  {\it all} the possible (physically distinct) embeddings in Einstein spaces, including the Ricci-flat ones widely discussed in the literature. We examine in detail their  generalization, which - in the framework  under consideration -    are  higher-dimensional spacetimes  sourced by a scalar field with flat (constant $\neq 0$)  potential.  We use the Kretschmann curvature scalar to show that many embedding spaces have  a physical  singularity at some finite value of the extra coordinate.
We develop several classes of embeddings that are free of singularities,  have distinct  non-vanishing self-interacting  potentials and   are continuously connected (in various limits) to Einstein embeddings. We point out that the induced metric possesses scaling symmetry and, as a consequence, the effective physical parameters (e.g., mass, angular momentum, cosmological constant) can be interpreted as functions of the extra coordinate.

\end{abstract}

\medskip


\newpage

\section{Introduction}

 Kaluza's great achievement was the discovery that extending the number of dimensions from four to five allows the unification of gravity and electromagnetism \cite{Kaluza}. The appearance of the ``extra" dimension in physical laws was avoided by imposing the ``cylinder condition", which essentially requires that all derivatives with respect to the fifth coordinate vanish. 

Modern higher-dimensional gravity theories, inspired by string theories, do not require this condition; it is replaced by the assumption that  matter fields are confined to our 4D spacetime, which is modeled as a singular hypersurface or ``brane" embedded in a larger $(4 + d)$ world, while gravity is a multidimensional interaction that can propagate in the extra $d$ dimensions as well \cite{Arkani1, Arkani2, Arkani3}.
Well-known examples of such brane theories are provided by the Randall-Sundrum \cite{RS1, RS2} and the Dvali-Gabadadze-Porrati scenarios \cite{Dvali1, Dvali2}. In the so-called space-time-matter theory, inspired by the unification of matter and geometry, the framework is similar; the fifth dimension is not compactified and  our spacetime is  identified with  some   four-dimensional hypersurface orthogonal to the extra dimension. The main difference is that in the latter the hypersurface is not necessarily singular and the effective matter in 4D is derived from vacuum in 5D, through the metric's dependence on the extra coordinate
 \cite{Wesson1, JPdeL1, JPdeLWesson, Wessonbook}.

Independently of the theoretical motivation, the existence of a putative large extra dimension offers a wealth of new physics. That is illustrated by  the effective equations for gravity in 4D, which predict five-dimensional local and non-local corrections to the usual general relativity in 4D \cite{Shiromizu, Maartens2, Chamblin, Maartens1, Maartens3, Maartens4, DadhichGosh, Dadhich2, DeruelleKatz, Saavedra}. Also, the geodesic equation for test particles in 5D predicts an effective four-dimensional equation of motion with an extra non-gravitational force \cite{MashhoonWesson, WessonMashhoon, Youm1, Youm2, Eq.ofMotion, JPdeLforce, Seahra3, JPdeL2}. As a consequence, there has been a growing interest in models where our  4D  spacetime is  embedded in a higher-dimensional space (for historical reviews and references see \cite{Pavsic, WessonHistory}), which in turn  has generated considerable attention to embedding theorems  of differential geometry and their applications to higher-dimensional theories of the universe.

In particular,  Campbell-Magaard's theorem \cite{Campbell}, which asserts that any  $n$-dimensional pseudo-Riemannian manifold may   be locally embedded in an $(n + 1)$- dimensional Ricci-flat space, implies that it is always
possible to locally embed any solution of the 4D Einstein equations of general relativity -  with an arbitrary energy-momentum tensor - in a Ricci-flat solution of the 5D vacuum Einstein equations. 
Motivated by the second Randall-Sundrum braneworld scenario and string theories, the mathematical proof of that theorem has been extended to include cases where
the higher-dimensional space has a nonzero cosmological
constant, is sourced by a scalar field, and
has an arbitrary non-degenerate Ricci tensor \cite{Rippl, Lidsey, AndersonLidsey, Dahia1, Dahia2, Seahra, Wessondefense}. However, it is
important to note two things: First, embedding theorems provide the conditions that  guarantee the existence of the embedding, but they do not show {\it how} to produce the actual embedding. Second, there are many ways of embedding a 4D spacetime in a higher-dimensional manifold  \cite{arXiv:gr-qc/0512067, JPdeLRussianJ}.

A neat example of this is provided by the problem of embedding vacuum solutions of $n$-dimensional general relativity in an $(n + 1)$-dimensional manifold. In fact, there are a number of works proposing specific embeddings of such solutions in Ricci-flat spaces, Einstein spaces, and spaces sourced by a  massless 
minimally-coupled scalar field. 
A natural question to ask is whether these (seemingly disconnected) embeddings are related to each other, and if so  how?
In this paper we present a systematic approach that allows us to tackle all these embeddings in a unified manner in such a way that we (i) reproduce known results, (ii) construct new embeddings and (iii) give concrete answers to the above-mentioned questions. Our results complement and generalize those obtained in \cite{MW1, MW2, AndersonLidsey, Anderson, Fonseca, Bejancu, MWL}. 

The paper is organized as follows. In Section $2$ we provide the  field equations and perform the   dimensional reduction. We show that there are infinite ways of building these embeddings as there are infinite ways of prescribing the embedding  function. In Section $3$ we construct  {\it all} the possible embeddings in Einstein spaces, including the Ricci-flat ones discussed in \cite{MW1, MW2}. By construction they are free of the degeneracy noted by  Fonseca-Neto and Romero \cite{Fonseca}. In Section $4$ we examine in detail the simplest generalization of Einstein  embeddings, which - in the framework  under consideration -    are spaces with one extra dimension    sourced by a scalar field with flat (constant $\neq 0$)  potential.  We use the Kretschmann curvature scalar, which we calculate in Appendix A, to show that many embedding spaces have  a physical  singularity at some finite value of the extra coordinate.
In Section $5$ we develop several classes of embeddings that are free of singularities,  have distinct  non-vanishing self-interacting  potentials and   are continuously connected (in various limits) to the Einstein embeddings of Section $3$. In Section $6$ we summarize some of the physical implications.  We point out that the induced metric possesses scaling symmetry and, as a consequence, the effective physical parameters (e.g., mass, angular momentum, cosmological constant) can be interpreted as  functions of the extra coordinate. We  show that  the non-vanishing logarithmic derivative  of the embedding function is responsible for the appearance of an extra force in $n$-dimensions.

\section{Field equations and dimensional reduction}

Let us consider an $n$-dimensional Lorentzian  manifold $(\Sigma, \, g_{a b})$, which is a solution of the  
$n$-dimensional Einstein
 equations of general 
relativity  in vacuum   
with cosmological constant $\Lambda_{(n)}$, i.e., 
\begin{equation}
\label{Field equations in n-dim}
{^{(n)}G}_{a b} = \Lambda_{(n)}\, g_{a b},\;\;\;\;(a, \, b = 0, \,1,\,...\, , \, n - 1).
\end{equation} 
Here, we will show    {\it{how}}   such a manifold  may be embedded in 
a  solution of the Einstein equations in $(n + 1)$ dimensions  
\begin{equation}
\label{Field equations in (n + 1)-dim}
{^{(n + 1)}G}_{A B} = \kappa\, {^{(n + 1)}T}_{A B},  \;\;\;\;(A, \, B = 0, \,1,\,...\, , \, n),
\end{equation}
where $\kappa$ represents  the gravitational constant in $(n + 1)$ dimensions and ${^{(n + 1)}T}_{A B}$, following Anderson {\it {et al.}} \cite{Anderson},  is the energy-momentum   
of a real massless scalar field $\Psi$, minimally-coupled to Einstein gravity and self-interacting through a potential $V(\Psi)$. The energy momentum tensor of such a field is 
\begin{equation}
\label{EMT for the scalar field}
{^{(n + 1)}T}_{A B} = \Psi_{A}\, \Psi_{B} - \frac{1}{2}\,\gamma_{A B}\,\Psi_{C} \Psi^C - \gamma_{A B}\, V(\Psi).
\end{equation}
Here $\Psi$ is dimensionless, $V$ has the units of (length)$^{- 2}$ and  $\gamma_{A B}$ is the metric of the $(n + 1)$-dimensional space. In this framework we will be able to expand and generalize a number of previous results  found in the literature. 

To construct an embedding for $(\Sigma, \, g_{a b})$ it is convenient to work in Gaussian normal coordinates where  the line element in $(n + 1)$ has the form 

\begin{eqnarray}
\label{metric in (n + 1)}
dS^2 = \gamma_{A B} \, d x^A d x^B &=& \gamma_{a b}\, d x^a d x^b + \epsilon d y^2,
\end{eqnarray}
where $y$ represents the coordinate along the $(n + 1)$-th dimension and  
$\epsilon = - 1$ or $\epsilon = 1$ depending on whether it is  spacelike or timelike, respectively. Next, 
 we adopt the ansatz used by Mashhoon and Wesson \cite{MW1, MW2}  and split $\gamma_{a b }$  into two parts via 
\begin{equation}
\label{split}
\gamma_{a b} = {\mbox{e}}^{2 \beta(x^{\rho}, y)}\;g_{a b}(x^{\rho}).
\end{equation}
Now we proceed to solve the field equations (\ref{Field equations in (n + 1)-dim})-(\ref{EMT for the scalar field}) subject to this split.

To compute the components of the Ricci tensor, we first calculate the Christoffel symbols. From (\ref{metric in (n + 1)})-(\ref{split}) it follows that $\gamma^{a b} = g^{a b}\, {\mbox{e}}^{- 2 \beta}$, $\gamma^{y y} = \gamma_{y y} = \epsilon$ and $\gamma^{y a} = 0$. Thus, we obtain 

\begin{eqnarray}
\label{Christoffel symbols, 1}
K_{y y}^{y} &=& K_{y y}^{a} = K_{a y}^{y} = 0\nonumber \\
K_{y a}^{b} &=& \beta'\, \delta_{a}^{b}, \;\;\;K_{a b}^{y} = - \epsilon \,\beta' \,{\mbox{e}}^{2 \beta}\,g_{a b}\nonumber\\
K_{a b}^{c} &=& \Gamma_{a b}^{c} + \left(\beta_{a}\,\delta_{b}^{c} + \beta_{b}\,\delta_{a}^{c} - \beta^{c}\,g_{a b}\right), 
\end{eqnarray}
where $K_{A B}^{C}$ and $\Gamma_{a b}^{c}$ are the Christoffel symbols formed from $\gamma_{A B}$ and $g_{a b}$, respectively; 
$\beta' = (\partial \beta/\partial y)$;  $\beta_{a} =  (\partial \beta/\partial x^a)$ and $\beta^a =  g^{a b}\,(\partial \beta/\partial x^b)$.

Consequently, the Ricci tensor for the metric (\ref{metric in (n + 1)})-(\ref{split}) can be written as
\begin{eqnarray}
\label{Ricci tensor in terms of Omega}
{^{(n + 1)}R}_{y a} &=& - (n - 1)\,\beta'_{a}\label{Ry a, in terms of beta}\\
{^{(n + 1)}R}_{y y} &=& - n\,(\beta'' + \beta'^2)\label{Ryy, in terms of beta}\\
{^{(n + 1)}R}_{a b} &=& {^{(n)}R}_{a b} + \sigma_{a b} - \epsilon\,\left(\beta'' + n \beta'^2\right)\,{\mbox{e}}^{2 \beta} \,g_{a b} \label{RAB in (n + 1), in terms of beta},
\end{eqnarray}
where
\begin{equation}
\label{definition of sigma}
\sigma_{a b} = (2 - n)\, \left[\beta_{a; b} - \beta_{a}\,\beta_{b}\right] - g_{a b}\,\left[(n - 2)\, \beta_{c}\,\beta^c + \beta_{c}^{c}\right].
\end{equation}
Here $\beta_{a; b}$ is the covariant derivative constructed with $\Gamma_{a b}^{c}$ and $\beta_{c}^{c} = g^{a b}\,\beta_{a; b}$.

The field equations (\ref{Field equations in (n + 1)-dim})-(\ref{EMT for the scalar field}) imply

\begin{equation}
\label{Field equations for Scalar field}
{^{(n + 1)}R}_{A B} = \kappa\,\left[\Psi_{A}\,\Psi_{B} + \frac{2 V}{(n - 1)}\, \gamma_{A B}\right].
\end{equation}
In what follows $n \neq 1$, because there is no Einstein gravity in one dimension (the concept of curvature requires at least two dimensions). 
Since $\gamma_{y a} = 0$, from this and (\ref{Ry a, in terms of beta}) it follows that $(n - 1)\,\beta'_{a} = - \kappa\, \Psi_{a}\, \Psi'$. This equation admits immediate integration if $\Psi = $ constant, or in a more general case when either $\Psi' = 0$ or  $\Psi_{a} = 0$. To make contact with \cite{MW1, MW2, Anderson}, we assume here that $\Psi_{a} = 0$, which means that $\Psi$ is either a constant or only depends on $y$. As a consequence ${^{(n + 1)}R}_{a y} = 0$ and we obtain

\begin{equation}
\label{separation of variables}
{\mbox{e}}^{2 \beta(x^{\rho}, y)} = F(y)\, {\mbox{e}}^{2 f(x^{\rho})},
\end{equation}
where $F(y)$ and $f(x^{\rho})$ are  arbitrary functions of integration.
In addition from (\ref{RAB in (n + 1), in terms of beta}) and (\ref{Field equations for Scalar field}) we get

\begin{equation}
\label{equation for Rab for scalar field}
{^{(n)}R}_{a b} = \left\{\frac{\epsilon}{2}\,\left[F'' + \frac{(n - 2)}{2}\,\frac{F'^2}{F}\right] + \frac{2\, \kappa\,V}{(n - 1)}\,F\right\}\, {\mbox{e}}^{2 f}\,g_{a b} - \sigma_{a b}.
\end{equation}
If $f = f_{0} = $ constant, then $\sigma_{a b} = 0$ and the embedded metric $g_{a b}$ represents an Einstein space. In this case, from (\ref{Field equations in n-dim}) it follows that the term multiplying the metric is $[2 \Lambda_{(n)}/(2 - n)]$. Thus we find

\begin{equation}
\label{equation for V, 1}
\kappa\,V = - \frac{\epsilon \,(n - 1)}{4}\,\left[\frac{F''}{F} + \frac{(n - 2)}{2}\, \frac{F'^2}{F^2} + \frac{4 \,\epsilon \tilde{\Lambda}_{(n)}}{(n - 2) F}\right],
\end{equation}
where $\tilde{\Lambda}_{(n)} = \Lambda_{n}\,{\mbox{e}}^{- 2 f_{0}}$. In what follows we omit the tilde.

Similarly, equating the expressions for  ${^{(n + 1)}R}_{y y}$ from (\ref{Ryy, in terms of beta}) and (\ref{Field equations for Scalar field}), and using (\ref{equation for V, 1}), we find

\begin{equation}
\label{equation for Psi' squared}
\kappa \Psi'^2  = - \frac{(n - 1)}{2}\,\left[\frac{F''}{F} - \frac{F'^2}{F^2}\right] + \frac{2\, \epsilon \Lambda_{(n)}}{(n - 2)\, F}. 
\end{equation}
Thus, the potential and the scalar field  are completely determined by the function $F(y)$.

Without loss of generality we can set $f_{0} = 0$, which is equivalent to a change of scale of coordinates, viz., ${\mbox{e}}^{ f_{0}}\,  x \rightarrow x$. After such a change, the line element in $(n + 1)$ becomes

\begin{eqnarray}
\label{metric in (n + 1), after change of scale}
dS^2 =  F(y) \,g_{a b} \, d x^a d x^b + \epsilon d y^2, 
\end{eqnarray}
which for every given $F$ provides an embedding  for $(\Sigma, \, g_{a b})$. The field equations do not give an equation for $F$, which reflects the fact that there are an infinite ways of embedding an $n$-dimensional 
manifold in an $(n + 1)$-dimensional space. 
To investigate the presence of singularities in these embeddings, in Appendix A we evaluate the Kretschmann scalar for metric (\ref{metric in (n + 1), after change of scale}). 

To complete the discussion we should mention that (\ref{equation for V, 1}) and (\ref{equation for Psi' squared}) are consistent with the equation of motion of the scalar field 
\begin{equation}
\label{equation for Psi}
\Psi^{C}_{;C} - \frac{d V}{d \Psi} = 0.
\end{equation}
In fact, for the line element (\ref{metric in (n + 1), after change of scale}) this equation reduces to 

\begin{equation}
\label{EMT for the scalar field, 1}
\Psi' \Psi'' + \frac{n}{2}\,\frac{F'}{F}\,\Psi'^2 - \epsilon V' = 0,
\end{equation}
 which is identically satisfied by (\ref{equation for V, 1}) and (\ref{equation for Psi' squared}). This is what we expected because  the equations of the gravitational field contain the equations for the matter which produces the field.

Also, we provide the non-vanishing mixed components of the energy-momentum tensor (\ref{EMT for the scalar field}). They are

\begin{eqnarray}
\kappa\, \, {^{(n + 1)}T}_{0}^{0} &=& \frac{\epsilon \, (n - 1)}{2}\,\left[\frac{F''}{F} + \frac{(n - 4)}{4}\,\frac{F'^2}{F^2}\right] + \frac{\Lambda_{(n)}}{F}\\
\kappa\, \, {^{(n + 1)}T}_{y}^{y} &=& \frac{\epsilon \, n (n - 1)}{8}\,\frac{F'^2}{F^2} + \frac{n\,\Lambda_{(n)}}{(n - 2)\,F},
\end{eqnarray}
and $T_{0}^{0} = T_{1}^{1} = \cdots = T_{n}^{n}\;\;\; \mbox{(no summation)}$. In what follows we set $\kappa = 1$.

In the above equations $n \neq 2$. However, this does not impose physical restrictions because in 
two-dimensions the Einstein tensor vanishes identically, so the cosmological constant $\Lambda_{(2)}$  must be zero. Therefore, if one wants to use these equations for $n = 2$ one should set $\left[\Lambda_{(n)}/(n - 2)\right]_{n = 2} = 0$ everywhere.

\subsection{Checking the equations}

 To check the above equations we apply them to the case where $F(y) = (1 + \lambda y)^{2 p}$ and $\Lambda_{(n)} = 0$ considered by Anderson {\it{et al.}} \cite{Anderson}. Here $p$ and $\lambda$ are constants.
Substituting into  (\ref{equation for Psi' squared}) we get 

\begin{equation}
\label{assumtion on F(y)}
\Psi'^2 = \frac{p \,(n - 1) \lambda^2}{(1 + \lambda y)^2}.
\end{equation}
Thus,

\begin{equation}
\label{Psi for Lambda = 0}
\Psi = q\, \ln|1 + \lambda y| + \Psi_{0},
\end{equation}
where $\Psi_{0}$ is a constant of integration and 
\begin{equation}
\label{q for Lambda = 0}
q^2 = p \, (n - 1)
\end{equation}
The corresponding potential (\ref{equation for V, 1}) is
\begin{equation}
\label{potential for Lambda = 0}
V(\Psi) = - \frac{\epsilon \, q^2\,\lambda^2\,(n p - 1)}{2}\, {\mbox{e}}^{- 2(\Psi - \Psi_{0})/q}.  
\end{equation}
Expressions (\ref{Psi for Lambda = 0})-(\ref{potential for Lambda = 0}) reproduce the results of section $3.2$  of \cite{Anderson} (in that paper $\epsilon = 1$).

\section{Einstein embeddings: $\Psi = $ constant}

The simplest way to construct a particular  embedding for the metric $g_{a b}$ is to prescribe the function $F(y)$. 
However, we have no physical arguments which would allow us to decide in favor of some  particular choice.

In this section we assume $\Psi = $ constant,  which means that the embedding universe is an Einstein space with, in general, nonzero Ricci curvature. This assumption provides an equation for $F$. Namely,  from (\ref{equation for Psi' squared}) we get

\begin{equation}
\label{equation for F, constant Psi. Version 1}
F F'' - F'^2 - \frac{4\, \epsilon \,\Lambda_{(n)}}{(n - 1) (n - 2)}\, F = 0.
\end{equation}
The first integral to this equation is
\begin{equation}
\label{First integral to equation for F, constant Psi. Version 1}
F'^2 = C\, F^2 - \frac{8 \,\epsilon \,\Lambda_{(n)}}{(n - 1) (n - 2)}\, F,  
\end{equation}
where $C$ is a constant of integration with dimensions of (length)$^{- 2}$. Substituting this into (\ref{equation for V, 1}) we find

\begin{equation}
\label{V in terms of C1}
V = - \frac{\epsilon \, C \, n (n - 1)}{8}. 
\end{equation}
In the case under consideration  the energy-momentum tensor (\ref{EMT for the scalar field}) reduces to ${^{(n + 1)}T}_{A B} = - V \gamma_{A B}$, which implies that $- V$ can be interpreted as the cosmological constant $\Lambda_{(n + 1)}$ in $(n + 1)$-dimensions. Thus, 
 \begin{equation}
\label{definition of C}
 C = \frac{8 \epsilon\, \Lambda_{(n + 1)}}{n (n - 1)}.
\end{equation} 
Consequently, when $\Psi = $ constant the function  $F$ is determined by the equation
\begin{equation}
\label{equation for F in terms of both Lambdas}
F'^2 = \frac{8\,\epsilon\,F}{n - 1}\left[\frac{\Lambda_{(n + 1)}}{n}\,F - \frac{\Lambda_{(n)}}{n - 2}\right].
\end{equation}

\subsection{The function $F(y)$ for Einstein embeddings}

Equation (\ref{equation for F in terms of both Lambdas}), for $n \neq 2$,  has seven distinct types of solutions that depend on $\Lambda_{(n)}$, $\Lambda_{(n + 1)}$ and $\epsilon$. For $n = 2$ it has a unique solution which is given by (\ref{solution for Lambda(n) = 0}) below.

\paragraph{Type I:} The simplest solution is $F = F_{0}$ = constant. From (\ref{equation for F, constant Psi. Version 1}) and (\ref{equation for F in terms of both Lambdas}) it follows that  $F$ is constant  only if $\Lambda_{(n)} = \Lambda_{(n + 1)} = 0$.

\paragraph{Type II:} These are  solutions for $\Lambda_{(n)} = 0$, $\Lambda_{(n + 1)} \neq 0$. From (\ref{equation for F in terms of both Lambdas}) we find that this case requires $\epsilon\,\Lambda_{(n + 1)} > 0$, which means that  the $(n + 1)$-dimensional  spacetime is  anti-de Sitter (de Sitter) if the extra dimension is spacelike (timelike). The solution is

\begin{equation}
\label{solution for Lambda(n) = 0}
F(y) = F_{0}\,{\mbox{e}}^{s \sqrt{C}\, y}, \;\;\;\;C > 0, \;\;s = \pm 1,  
\end{equation}
where $F_{0}$ is a constant of integration and $C$ is given by (\ref{definition of C}). 
When $\Lambda_{(n + 1)} \rightarrow 0$ we recover the constant solution.

This type of embedding is reminiscent of the reduction ansatz used in the  RS-2 scenario of braneworld theory\footnote{In that scenario  $y$ in (\ref{solution for Lambda(n) = 0}) is replaced by its modulus, so that $F'$ is discontinuous at $y = 0$, which gives a delta-function contribution to ${^{(n + 1)}R}_{A B}$.} \cite{RS2}. It allows to embed any solution of the $n$-dimensional  Einstein field equations in vacuum with 
zero cosmological constant in an $(n + 1)$-dimensional Einstein space with a nonzero cosmological constant.

\paragraph{Type III:} These are  solutions for $\Lambda_{(n + 1)} = 0$, $\Lambda_{(n)} \neq  0$. In this case the   $(n + 1)$-spacetime is usually called Ricci-flat. To obtain solutions to 
(\ref{equation for F in terms of both Lambdas}) in terms of real functions we should  take $\epsilon \Lambda_{(n)} < 0$. Also, to simplify the notation we introduce the quantity

\begin{equation}
\label{the quantity}
            \bar{\Lambda}_{(n)} =  \frac{6 \Lambda_{(n)}}{(n - 1) (n - 2)},
\end{equation}
in terms of which the solution can be written as

\begin{equation}
\label{appropriate form of the solution for C = 0}
F(y) = \left[\sqrt{\frac{- \epsilon \, \bar{\Lambda}_{(n)}}{3}} \; y + \sqrt{F_{0}}\right]^2.
\end{equation}
It reduces to  the constant solution  when $\Lambda_{(n)} \rightarrow 0$. This embedding generalizes to an arbitrary number of dimensions the ones obtained  in \cite{MW1, MW2, Fonseca}.

\medskip

When $\Lambda_{(n)} \neq 0$ and $\Lambda_{(n + 1)} \neq 0$ the solutions of (\ref{equation for F in terms of both Lambdas}) crucially depend on the sign   of $\epsilon \,\Lambda_{(n + 1)}$. If  $\epsilon\, \Lambda_{(n + 1)} >  0$, then    
$\epsilon\, \Lambda_{(n)}$ can be positive, negative or zero and the solutions are expressed in terms of exponentials.  If $\epsilon\, \Lambda_{(n + 1)} < 0$, then 
$\epsilon  \bar{\Lambda}_{(n)}$ must be negative and the solutions are oscillatory. 
In addition, their behavior  depends on the choice of the remaining constant of integration.  Here we choose it in such a way that we  recover the solutions of Type I - Type III in the appropriate limits.

\paragraph{Type IV:} These are solutions for $\epsilon\, \Lambda_{(n + 1)} > 0$ 
and $\Lambda_{(n)} \neq 0$, which  contain those of Type II as a limiting case $(s\,\sqrt{C}\, y \gg 1)$. There are three ``subtypes".

\medskip

$\bullet$ {\bf {Type IVa}}. These are embeddings that admit
both signs of $\epsilon\, \bar{\Lambda}_{(n)}$. In the limit $\Lambda_{(n)} \rightarrow 0$ they yield  (\ref{solution for Lambda(n) = 0}), independently of $\Lambda_{(n + 1)}$ (or $C$). Namely,

\begin{equation}
\label{Solution Type IVa}
F(y) = F_{0}\, {\mbox{e}}^{s\, \sqrt{C}\, y}\, \left[1 + \frac{\epsilon\, {\bar{\Lambda}}_{(n)}}{3 F_{0} \,C}\, {\mbox{e}}^{- s\, \sqrt{C}\, y}\right]^2, \;\;\;\;\;C \neq 0.
\end{equation}

\medskip

$\bullet$ {\bf {Type IVb}}.   
These  embeddings require  $\epsilon\, \bar{\Lambda}_{(n)} < 0$, and in the limit $\Lambda_{(n + 1)} \rightarrow 0$ reproduce (\ref{appropriate form of the solution for C = 0}), viz.,  

\begin{equation}
\label{Solution Type IVb}
F(y) = - \frac{4 \,\epsilon \,\bar{\Lambda}_{(n)}}{3 C}\, \sinh^2\left[\frac{\sqrt{C}\, (y + \tilde{y})}{2}\right], \;\;\;\;\epsilon \,\bar{\Lambda}_{(n)} < 0,
\end{equation}
where $\tilde{y}$ is a constant. If we choose $\tilde{y} = \left({- 3\, \epsilon\, F_{0}/\bar{\Lambda}_{n}}\right)^{1/2}$, then in the limit $C \rightarrow 0$ ($\Lambda_{(n + 1)} \rightarrow 0$) we recover the Ricci-flat embedding (\ref{appropriate form of the solution for C = 0}). When $\epsilon = - 1$ this function allows to embed any solution of the $n$-dimensional  Einstein field equations in vacuum with $\Lambda_{(n)} > 0$
 in an $(n + 1)$-dimensional Einstein space with $\Lambda_{(n + 1)} <  0$.

  $\bullet$ {\bf {Type IVc}}. These embeddings  require $\epsilon\, \bar{\Lambda}_{(n)} > 0$ and do not support the limits $\Lambda_{(n + 1)} \rightarrow 0$, $\Lambda_{(n)} \rightarrow 0$, separately. Namely,   
\begin{equation}
\label{Solution Type IVa, 2}
F(y) = \frac{4 \,\epsilon \,\bar{\Lambda}_{(n)}}{3 C}\, \cosh^2\left[\frac{\sqrt{C}\, (y + \tilde{y})}{2}\right], \;\;\;\;\epsilon \bar{\Lambda}_{(n)} > 0.
\end{equation}
However, if we assume that the cosmological constants are related, e.g., as
\begin{equation}
\label{relation between the Lambdas}
\Lambda_{(n)} =   F_{0}\; \frac{(n - 2)}{n}\,\Lambda_{(n + 1)},\;\;\;\;\;F_{0} = \mbox{constant $> 0$}
\end{equation}
 then (\ref{Solution Type IVa, 2}) becomes (without loss of generality we set $\tilde{y} = 0$)
\begin{equation}
\label{Solution Type IVa, 3}
F(y) = F_{0} \,  \cosh^2\left(\frac{\sqrt{C}\, y}{2}\right), \;\;\;\;\epsilon \bar{\Lambda}_{(n)} > 0.
\end{equation}
 which in the limit  $\Lambda_{(n)} \rightarrow 0$ ($\Lambda_{(n + 1)} \rightarrow 0$)  is well-behaved  and reproduces the constant (Type I) solution. When $\epsilon = - 1$ this function allows to embed any solution of the $n$-dimensional  Einstein field equations in vacuum with $\Lambda_{(n)} <  0$
 in an $(n + 1)$-dimensional Einstein space with $\Lambda_{(n + 1)} < 0$. By virtue of the relation (\ref{relation between the Lambdas}), here $\Lambda_{(n)}$ is  an adjustable parameter; by selecting $F_{0}$ appropriately it can be chosen to be
small enough to be phenomenologically realistic,  regardless of the size of $\Lambda_{(n + 1)}$ \cite{Pope}.

\paragraph{Type V:} These are the trigonometric counterparts of (\ref{Solution Type IVb}), (\ref{Solution Type IVa, 2}) and  (\ref{Solution Type IVa, 3}), which are the solutions   to (\ref{equation for F in terms of both Lambdas})  when   $\epsilon\, \Lambda_{(n + 1)} < 0$. All of them 
require  $\epsilon  \bar{\Lambda}_{(n)} < 0$, since $F$ is nonnegative. 

 The counterpart of (\ref{Solution Type IVb}) is a new solution, viz.,

\begin{equation}
\label{Solution Type Va}
F(y) = - \frac{4 \,\epsilon \,\bar{\Lambda}_{(n)}}{3 \omega^2}\, \sin^2\left[\frac{\omega\, (y + \tilde{y})}{2}\right],  \;\;\;\;\omega = \sqrt{- C} > 0, \;\;\;\epsilon \, \bar{\Lambda}_{(n)} < 0.
\end{equation}
In the limit $C \rightarrow 0$ ($\Lambda_{(n + 1)} \rightarrow 0$) it reduces to  the Ricci-flat embedding (\ref{appropriate form of the solution for C = 0}). 
However, the trigonometric counterpart of (\ref{Solution Type IVa, 2}), with $C = - \omega^2$,  is not a  new solution because after a   simple change $\tilde{y} \rightarrow (\tilde{y} - \pi/\omega)$ it reduces to (\ref{Solution Type Va}). If we apply the same procedure to (\ref{Solution Type IVa, 3}) we find that 

\begin{equation}
\label{Solution Type Vb}
F(y) = F_{0}\, \cos^2\left(\frac{\omega\,y }{2}\right),  \;\;\;\;\omega = \sqrt{- C},
\end{equation}
is an embedding function  only if (\ref{relation between the Lambdas}) holds true, viz.,  $\omega = \sqrt{- C} = \sqrt{ - \frac{4\, \epsilon \bar{\Lambda}_{(n)}}{3 F_{0}}}$. 
For $\epsilon = - 1$,  (\ref{Solution Type Va}) and (\ref{Solution Type Vb}) serve to embed  $n$-dimensional vacuum solutions with  $\Lambda_{(n)} > 0$
 in an $(n + 1)$-dimensional Einstein space with $\Lambda_{(n + 1)} >  0$. The difference is that  $\Lambda_{(n)}$ and $\Lambda_{(n + 1)}$ are independent parameters in (\ref{Solution Type Va}), but not in (\ref{Solution Type Vb}).

\subsection{Singularities of the Einstein embeddings}

To investigate the singularities we use the Kretschmann curvature scalar. Substituting (\ref{equation for F, constant Psi. Version 1})-(\ref{First integral to equation for F, constant Psi. Version 1}) into (\ref{Kretschmann scalar in (n + 1), explicit expression}) we obtain

\begin{equation}
\label{Kretschmann scalar in (n + 1). Einstein embeddings}
\widehat{K}^2 = \frac{1}{F^2}\left[ K^2 - \frac{2}{9}\, n (n - 1)\, \bar{\Lambda}_{(n)}^2\right] + \frac{8\,(n + 1)}{n (n - 1)^2}\, \Lambda_{(n + 1)}^2.
\end{equation}
Note that for embeddings of Type II, $F$   becomes zero only asymptotically, as $(s \,y) \rightarrow - \infty$. Those of Type IV with $\epsilon \Lambda_{(n)} > 0$ have $F(y)  > 0$,  for all values of $y$.  However,  $F$ does reach zero at some finite value of $y$, say $y = y_{*}$,  when $\epsilon \Lambda_{(n)} < 0$. In the latter case, $y = y_{*}$ is a curvature singularity, unless the term in square bracket is zero. As far as we know, the only case where this is so is the n-dimensional de 
Sitter space.\footnote{In fact, the Riemann curvature tensor of de Sitter space in $n$-dimensions is given by ${^{(n)}R}_{a b c d} = \frac{2\,\Lambda_{(n)}}{(n - 1) (n - 2)}\,\left(g_{a d}\, g_{b c} - g_{a c}\, g_{b d} \right)$. In terms of $\bar{\Lambda}_{(n)}$,  introduced in (\ref{the quantity}),  we obtain $K^2 = \frac{2}{9}\, n (n - 1)\, \bar{\Lambda}_{(n)}^2$.} So the embeddings of\footnote{dS$_{n} \equiv$  de Sitter spacetime in $n$-dimensions;  AdS$_{n} \equiv $ Anti-de Sitter spacetime in $n$-dimensions.} dS$_{n}$ or AdS$_{n}$ in dS$_{n + 1}$ or 
AdS$_{n + 1}$ are the only ones which are free of singularities at $y_{*}$; the embedding of any other metric, for example the 
Schwarzschild-de Sitter metric, will be singular at those hypersurfaces.

It may be verified that  the behavior of all solutions with $\epsilon\, \Lambda_{(n)} < 0$, namely (\ref{appropriate form of the solution for C = 0})-(\ref{Solution Type IVb}) and (\ref{Solution Type Va})-(\ref{Solution Type Vb}),  near $y_{*}$ is\begin{equation}
F(y)\,  \stackrel{y \rightarrow y_{*}}{\longrightarrow} -\frac{\epsilon \, \bar{\Lambda}_{(n)}}{3}\, (y - y_{*})^2.
\end{equation}
With the transformation $\tilde{y} = (y - y_{*})$, which is a  change to a new zero point for $y$, the line element (\ref{metric in (n + 1), after change of scale}) near   $y_{*}$   becomes (omitting the tilde)

\begin{equation}
\label{asymptotic form of the Einstein embeddings}
d S^2 \,  \stackrel{y \rightarrow 0}{\longrightarrow} \;- \frac{\epsilon \, \bar{\Lambda}_{(n)}}{3}\, y^2 \, g_{a b} \,d x^a\, d x^b + \epsilon\, d y^2, \;\;\;\;\;\epsilon \, \bar{\Lambda}_{(n)} <0. 
\end{equation}
When $n = 4$ and $\epsilon = - 1$ this reduces to  the (pure) canonical metric  employed by Mashhoon and Wesson \cite{MW1, MW2} to embed vacuum solutions with $\Lambda_{(4)} > 0$ in a five-dimensional Ricci-flat space. Consequently, we can assert  that  the  (pure) canonical metric in $(n + 1)$-dimensions (\ref{asymptotic form of the Einstein embeddings}) describes the asymptotic behavior of Einstein embeddings, with $\Lambda_{(n + 1)} \neq 0$, near the hypersurface $y = 0$, which - except for dS$_{n}$ and AdS$_{n}$ - is a singular one for all embeddings of vacuum solutions of $n$-dimensional general relativity.

\section{Flat potential: $V(\Psi) = V_{0} = $ constant}

This is the simplest  generalization of the Einstein embeddings discussed in the previous section. In fact,  
when $V = V_{0} = $ constant, (\ref{equation for V, 1}) provides  and equation for $F$

\begin{equation}
\label{equation for F. Vanishing potential}
F F'' +  \frac{(n - 2)}{2}\, F'^2 + \frac{4 \epsilon \Lambda_{(n)}}{(n - 2)}\, F  + \frac{4 \, \epsilon V_{0}\,F^2}{(n - 1)}= 0,
\end{equation}
whose  first integral  is 

\begin{equation}
\label{First integral, equation for F. Vanishing potential}
F'^2 =   \frac{E}{F^{(n - 2)}}  - \frac{8 \epsilon \,\Lambda_{(n)}\, F}{(n - 1) (n - 2)}   - \frac{8 \epsilon\, V_{0}\, F^2}{n (n - 1)},
\end{equation}
where $E$ is a constant of integration with dimensions of (length)$^{-2}$.
Inserting this into (\ref{equation for Psi' squared}) we find
\begin{equation}
\label{equation for Psi when V = const}
\Psi'^2 = \frac{E \, n (n - 1)}{4 F^{n}},\;\;\;\;E \geq 0, 
\end{equation}
which is consistent with (\ref{EMT for the scalar field, 1}). Note that $\Psi =$ constant when $E = 0$, and (\ref{First integral, equation for F. Vanishing potential})  becomes identical to (\ref{equation for F in terms of both Lambdas}) with a cosmological constant in the $(n + 1)$-dimensional space given by $\Lambda_{(n + 1)} = - V_{0}$, as expected. Thus, the Einstein embeddings correspond to $E = 0$.

Unfortunately there is no a general algebraic solution to  these equations, but  we can construct   a mechanical analog which helps us to understand the nature of the solutions. To this end we set
\begin{equation}
\label{F in term s of X} 
F = X^{2/n}, 
\end{equation}
which allows to write (\ref{First integral, equation for F. Vanishing potential}) as  

\begin{equation}
\label{mechanical equivalent}
\frac{1}{2}\, \mu\,  X'^2 = E - U(X),  
\end{equation}
where $\mu = (8/n^2)$ and 
\begin{equation}
\label{mechanical potential}
U(X) = \frac{8\, \epsilon }{n - 1}\,\left[\frac{\Lambda_{(n)}}{(n - 2)}\, X^{2 (n - 1)/n} + \frac{V_{0}}{n }\, X^2\right],
\end{equation}
which in the language of classical mechanics describes a particle of mass $\mu$ moving in the $X$-direction, under the action of a $X$-directed conservative force with potential $U(X)$, and total constant energy $E$.

\subsection{The embedding function $F(y)$ for flat potential}

Once again, there are seven distinct types of solutions depending on $\epsilon \, V_{0}$ and $\epsilon \, \Lambda_{(n)}$.

\paragraph{1) Solutions for $\Lambda_{(n)} \neq 0$ and $F = F_{0}$ constant:} This is the simplest solution to (\ref{equation for F. Vanishing potential})-(\ref{equation for Psi when V = const}). It is given by   

\begin{eqnarray}
\Psi(y) &=& \sqrt{\frac{2 \epsilon \, \Lambda_{(n)}}{(n - 2)\, F_{0}}} \; y + \Psi_{0}\label{Psi for F constant}, \;\;\;\;\;\epsilon \, \Lambda_{(n)} > 0\\
V_{0} &=& - \frac{(n - 1)\, \Lambda_{(n)}}{(n - 2)\, F_{0}}\label{V for F constant},\;\;\;\;\frac{\Lambda_{(n)}}{V_{0}} < 0\\
E &=& \frac{4 \,\epsilon\,{\bar{\Lambda}}_{{(n)}}}{3\, n}\, \left[\left(\frac{n - 1}{n - 2}\right)\,\left(- \frac{\Lambda_{(n)}}{V_{0}}\right)\right]^{(n - 1)}\label{E for F constant},
\end{eqnarray}
where to simplify the notation we have used (\ref{the quantity}). They describe the unstable equilibrium position at the top of the curve $U(X)$, see solutions 7 below.  
 In the limit  $\Lambda_{(n)} \rightarrow 0$  we recover the solutions  of Type I.

\paragraph{2) Solutions for $\Lambda_{(n)}$ = 0 and $V_{0} = 0$:} In this case $U = 0$, which  describes a mechanical system in inertial motion. The solution in terms of the field quantities is

\begin{eqnarray}
F(y) &=& \left|\frac{n\,\sqrt{E}}{2} \, y + 1\right|^{2/n}\label{F for Lambda = 0, V = 0}\\
\Psi(y) &=& \pm\, \sqrt{\frac{n - 1}{n}}\, \ln \left|\frac{n\,\sqrt{E}}{2} \, y + 1\right| + \Psi_{0}\label{Psi for Lambda = 0,  V  = 0}.
\end{eqnarray}
For $\frac{n\,\sqrt{E}}{2} = \lambda$  the solution  is identical to that of Anderson {\it {et al.}}  \cite{Anderson}, given by  (\ref{Psi for Lambda = 0}),  with $p = 1/n$ for which the potential (\ref{potential for Lambda = 0}) vanishes. Changing the origin $y \rightarrow (y - 2/n \sqrt{E})$ it can be written as 

\begin{equation}
\label{solution 2 after changing the origin}
F(y) = \left(\frac{n\, \sqrt{E}}{2}\, y\right)^{2/n}, \;\;\;\;\Psi(y) = \pm \sqrt{\frac{n - 1}{n}}\, \ln \left(\frac{n\, \sqrt{E}}{2}\, |y|\right).
\end{equation}

\paragraph{3) Solutions for $\Lambda_{(n)}$ = 0 and $\epsilon\, V_{0} >  0$:} Now $U$ is positive and quadratic in $X$. Therefore,  (\ref{mechanical equivalent})-(\ref{mechanical potential}) represent a simple harmonic oscillator.  The corresponding solution for $F$  and $\Psi$ is given by 

\begin{eqnarray}
F(y) &=& \left[\frac{n \,\sqrt{E}}{2\, \omega}\, \sin \left(\omega \, y + \phi\right)\right]^{2/n}, \;\;\;\;\omega =  \sqrt{\frac{2 \, n \,\epsilon\,V_{0}}{n - 1}}\label{F for Lambda = 0, epsilon V > 0}\\
\Psi(y) &=&   \sqrt{\frac{n - 1}{n}}\,\ln \tan \left(\frac{\omega\, y + \phi}{2}\right) + \Psi_{0}\label{Psi for Lambda = 0, epsilon V > 0},
\end{eqnarray}
where $\phi$ is a constant of integration.    Here the extra dimension  $y$ is restricted to the range $0 < (\omega\,  y + \phi) <  \pi$.

\paragraph{4) Solutions for $\Lambda_{(n)}$ = 0 and $\epsilon\, V_{0} < 0$:} Again $U$ is quadratic in $X$,  but this time it is negative. Therefore the  functions $F$ and $\Psi$ have the same mathematical shape as in (\ref{F for Lambda = 0, epsilon V > 0})-(\ref{Psi for Lambda = 0, epsilon V > 0}), except that now the trigonometric  functions are replaced by the corresponding hyperbolic functions and 
   $\omega = \sqrt{- 2\, n \epsilon\, V_{0}/(n - 1)}$. The physical consequence is that  the extra dimension $y$ can take arbitrary large values. 
In such a case,   $F(y) \rightarrow \mbox{e}^{2 \omega y/n}$  and  $\Psi' \rightarrow 0$ which implies ${^{(n + 1)}T}_{A B} \rightarrow - \gamma_{A B}\, V_{0} = \gamma_{A B}\, \Lambda_{{(n + 1)}}$.  
Then, by virtue of the relation between $\omega$ and $V_{0}$, for large values of $(\omega \, y + \phi)$ these solutions approach those of Type II discussed in the previous section. 

\medskip

$\bullet$ If we set $\phi = (2 \, \omega/ n\, \sqrt{E})$, then  in the limit $V_{0} \rightarrow 0$ the last two solutions  exactly reproduce (\ref{F for Lambda = 0, V = 0})-(\ref{Psi for Lambda = 0, V = 0}).

\paragraph{5) Solutions for $V_{0}$ = 0 and $\epsilon\,\Lambda_{(n)} < 0$:} In this case  $U(X) < 0$ and the motion is unbounded, i.e.,  $0 < X < \infty$. The solutions cannot be expressed in terms of elementary functions, except for $E = 0$. But, their qualitative behavior is as follows: Near the origin 
they are  well represented  by (\ref{solution 2 after changing the origin}). For $X \gg 1$ they  mimic those of Type III (\ref{appropriate form of the solution for C = 0}) for which $\Lambda_{(n + 1)} = 0$.

\paragraph{6) Solutions for $V_{0}$ = 0 and $\epsilon\,\Lambda_{(n)} > 0$:} Now $U(X) > 0$, the (fictitious) particle moves under the action of a nonlinear restoring force $\ \sim -  X^{(n - 2)/n}$, and $X$ is bounded to move between zero and $X_{max}$, which is given by the solution of the equation $E = U(X)$. In terms of $F$ we have $0 \leq F \leq \left(\frac{3 E}{4\,\epsilon \bar{\Lambda}_{(n)}}\right)^{1/(n - 1)}$.

\paragraph{7) Solutions for $V_{0} \neq 0$ and $\Lambda_{(n)} \neq 0$:} We introduce the dimensionless parameter $\alpha$ as 
\begin{equation}
\label{definition of alpha}
\frac{V_{0}}{n} = \alpha\, \frac{\Lambda_{(n)}}{(n - 2)}.
\end{equation}
Then, using the quantity ${\bar{\Lambda}}_{{(n)}}$ introduced in (\ref{the quantity}), the function $U$ becomes
\[
U(X) = \frac{4}{3}\, \epsilon\,{\bar{\Lambda}}_{{(n)}}\,\left(X^{2(n - 1)/n} + \alpha X^2\right). 
\]
There are different solutions depending on the sign of $\epsilon \, \Lambda_{(n)}$ and $\alpha$. 

Perhaps the most interesting are those with $\epsilon \, \Lambda_{(n)} > 0$ and $\alpha < 0$. In that case $U$ vanishes not only at $X = 0$ but also at $X = (-1/\alpha)^{n/2}$. Thus,  $U$ has a maximum $U_{0} = \frac{4 \,\epsilon\,{\bar{\Lambda}}_{{(n)}}}{3\, n}\, \left[\frac{(n - 1)}{( - \alpha)\, n}\right]^{(n - 1)}$ at $X_{0} = \left[\frac{n - 1}{(- \alpha)\,n}\right]^{n/2}$. 
Consequently, for $E < U_{0}$ the motion is bounded. For $E > U_{0}$, $X$ (or $F$) can take arbitrary large values. Asymptotically, for $F \gg 1$ the solutions coincide with those of Type II. 
For $E = U_{0}$, the system eventually reaches the equilibrium point $X_{0}$ where $X' = X'' = 0$, which  corresponds to a state of unstable equilibrium, since $U(X)$ has a maximum at that point. Such a state is described by (\ref{Psi for F constant})-(\ref{E for F constant}).

 For $\epsilon \, \Lambda_{(n)} < 0$ and $\alpha < 0$ the solutions are oscillatory and generalize those of Type V to the case where $E > 0$.

\subsection{Singularities of the embeddings   with flat potential}

To obtain the Kretschmann scalar for these embeddings we substitute (\ref{equation for F. Vanishing potential})-(\ref{First integral, equation for F. Vanishing potential}) into  (\ref{Kretschmann scalar in (n + 1), explicit expression}) and get

\begin{equation}
\label{Kretschmann scalar for embeddings with flat potential}
\widehat{K}^2 = \frac{1}{F^2}\, \left[K^2 - \frac{2}{9}\, n (n - 1)\, \bar{\Lambda}_{(n)}^2\right] + \frac{8 (n + 1)}{n (n - 1)^2}\, V_{0}^2 + \frac{2\, E}{F^n}\, \left[\epsilon\, V_{0} + \frac{E \, (2 n - 1)(n - 1) n}{16 \, F^n}\right].
\end{equation}
For $E = 0$ we recover (\ref{Kretschmann scalar in (n + 1). Einstein embeddings}), as expected. Except for (\ref{Psi for F constant})-(\ref{E for F constant}), all solutions of (\ref{equation for F. Vanishing potential})-(\ref{First integral, equation for F. Vanishing potential}) with $E > 0$ become zero at some finite $y = y_{*}$, which is a curvature singularity even when  the embedded spacetime is dS$_{n}$ or AdS$_{n}$, and the first square bracket vanishes. We can choose the origin of $y$ to make $F(0) = 0$. With this choice, the behavior of the solutions near the singularity is given by (\ref{solution 2 after changing the origin}) and the line element is 

\begin{equation}
d S^2 \,  \stackrel{y \rightarrow 0}{\longrightarrow} \; \left(\frac{n \, \sqrt{E}}{2}\, y\right)^{2/n} \, g_{a b}\,d x^a\, d x^b + \epsilon\, d y^2, 
\end{equation}
regardless of $\Lambda_{(n)}$ and $V_{0}$.

\section{General case: $\Psi \neq$ constant, $V \neq$ constant}

As mentioned earlier, we have a system of two equations, namely (\ref{equation for V, 1}) and (\ref{equation for Psi' squared}),  for  three unknown functions, $(F,\,\Psi, \, V)$. One way to close the system is to prescribe a function $V = V(\Psi)$. However, in absence of such a function,  a more appealing  procedure  is to  assume either  $\Psi = $ constant or $V = $ constant, which is what we have done in the last two Sections. Otherwise, there are  infinite  ways of prescribing the embedding function  $F$.

In this section, we build several examples.  To reduce the arbitrariness we select  $ F $ in such a way that, in the appropriate limits, we regain some of the previously discussed embedding spacetimes.

\paragraph{Example I:} As a model for the choice of $F$ we use (\ref{solution for Lambda(n) = 0}). Thus we assume

\begin{equation}
\label{F for example I}
F = F_{0}\,{\mbox{e}}^{\lambda \,y},
\end{equation}
where $\lambda$ is a parameter with dimensions of (length)$^{ - 1}$. In this case $F \neq 0$ for any finite value of $y$ and the Kretschmann scalar is
\begin{equation}
\label{Kretschmann for Model I}
{\widehat{K}}^2 = \frac{K^2}{F^2} + \frac{2 n \,\epsilon \,\Lambda_{(n)}\, \lambda^2}{(n - 2)\, F}+ \frac{n (n + 1)\, \lambda^4}{8}.
\end{equation}
Thus,  there are no singularities in $(n + 1)$ dimensions, except for those which come from the embedded spacetime.
The corresponding scalar field and it potential can be written as

\begin{eqnarray}
\Psi(y) &=& \frac{2}{\lambda}\,\sqrt{\frac{2\, \epsilon\, \Lambda_{(n)}}{(n - 2) \, F_{0}}}\, \left(1 - \mbox{e}^{- \lambda y/2}\right) + \Psi_{0} \label{Psi for F exponential}\\
V(\Psi) &=& - \frac{(n - 1)\, \Lambda_{(n)}}{(n - 2)\, F_{0}} + \frac{\epsilon\, \lambda (n - 1)}{2}\,\left[\sqrt{\frac{2\, \epsilon\, \Lambda_{(n)}}{(n - 2) \, F_{0}}}\, \left(\Psi - \Psi_{0}\right) - \frac{\lambda}{4}[n + \left(\Psi - \Psi_{0}\right)^2]\right]\label{V for F exponential}.
\end{eqnarray}

$\bullet$ In the limit $\lambda \rightarrow 0$ these expressions reduce to (\ref{Psi for F constant}) and (\ref{V for F constant}), respectively.

\medskip

$\bullet$ If $\Lambda_{(n)} = 0$, then $\Psi = \Psi_{0}$ and  $V$ is constant. In this case the  energy-momentum tensor (\ref{EMT for the scalar field}) reduces to
\begin{equation}
\label{EMT for F exponential}
{^{(n + 1)}T}_{A B} = \frac{\epsilon \,{n (n - 1)\,\lambda^2}}{8}\, \gamma_{A B}.
\end{equation}
 Thus, when $\Lambda_{(n)} = 0$ the embedding space is an Einstein space with an effective cosmological constant
\begin{equation}
\label{Cosmological constant in (n + 1) dim}
\Lambda_{(n + 1)} = \frac{\epsilon \,{n (n - 1)\,\lambda^2}}{8},
\end{equation}
which implies $\lambda = \pm \,\sqrt{C}$, where $C$ is given by (\ref{definition of C}).

 \paragraph{Example II:} Motivated by (\ref{Solution Type IVa}) we now choose the functional form
\begin{equation}
\label{functional form 1}
F(y) = \frac{F_{0}}{(1 + b)^2}\, \mbox{e}^{\lambda y}\, \left(1 + b\,  \mbox{e}^{- \lambda y}\right)^2,\;\;\;\;b \geq 0,
\end{equation}
which for $\lambda  \rightarrow 0$ reduces to the case $F = F_{0}$ given by  (\ref{Psi for F constant})-(\ref{V for F constant}),  and for $b \rightarrow 0$ gives back (\ref{F for example I})-(\ref{V for F exponential}).
In fact, substituting this into (\ref{equation for Psi' squared}) and choosing the integration constant appropriately we get  

\begin{eqnarray}
\label{field for IVa}
\Psi (y) &=& - \frac{2 A}{\lambda\,\sqrt{b}}\; \left[\arctan{(\sqrt{b}\, \mbox{e}^{- \lambda y/2})} - \arctan{(\sqrt{b})}\right] + \Psi_{0}
\end{eqnarray}
where 
\begin{equation}
\label{definition of A}
A^2 = \frac{2\; \epsilon \, \Lambda_{(n)}\, (1 + b)^2}{(n - 2) \, F_{0}} - \lambda^2\, b \,(n - 1).
\end{equation}
Clearly this equation requires  $\epsilon \Lambda_{(n)} > 0$. The corresponding potential can be written as

\begin{equation}
\label{potential for IVa}
V(\Psi) = - \frac{\epsilon}{8 \, b}\, (n - 1)\, \left[A^2 \,\sin^2 \alpha(\Psi) + n \lambda^2 b\right], 
\end{equation}
with 
\begin{equation}
\label{definition of alpha}
\alpha(\Psi) = \frac{\lambda\, \sqrt{b}}{A}\, \left(\Psi - \Psi_{0}\right) - 2 \arctan (\sqrt{b}).
 \end{equation}

$\bullet$ For $A = 0$, the field and the potential are constant. In this case the energy-momentum tensor in $(n + 1)$ dimension reduces to (\ref{EMT for F exponential}) and $\lambda = \pm \sqrt{C}$, where $C$ is given by (\ref{definition of C}). Thus, $A = 0$ implies

\[
b = \frac{\epsilon \, {\bar{\Lambda}}_{(n)}}{3\,C}\,\frac{(1 + b)^2}{F_{0}}
\]
After a rescaling $F_{0} \rightarrow F_{0}\,(1 + b)^2$, the function $F$ given by (\ref{functional form 1}) becomes identical to (\ref{Solution Type IVa}).

\medskip

Thus, the function $F$ does not determine, in a unique way, the energy-momentum tensor in $(n + 1)$-dimensions; choosing $F$ in such a way that it has the same functional form as in  an Einstein embedding,  we can generate a new family of embeddings, with nontrivial $\Psi$ and $V$, which contains the original one as a limiting case.  This is something one should have expected a priori, because  in general relativity  the same geometry can be engendered by different material distributions.\footnote{A nice example  is provided by the  FRW dust models which can be interpreted as exact solutions of the Einstein field equations for a viscous fluid  \cite{ColeyTupper}.}

\paragraph{Example III:} As a final example we consider the embedding of vacuum solutions of $n$-dimensional general relativity with $\Lambda_{(n)} = 0$. In this case (\ref{equation for Psi' squared}) can be formally integrated as   
\begin{equation}
\label{Function F, example III}
F(y) = F_{0} \, \mbox{e}^{a \,y}\, \mbox{e}^{- \int f(y)\, d y}, 
\end{equation}
where $F_{0}$ and $a$ are constants of integration and $f(y)$ is given by

\begin{equation}
\label{Psi prime for example III}
\Psi'^2 = \frac{(n - 1)}{2}\, f'.
\end{equation}
Substituting (\ref{Function F, example III}) into (\ref{equation for V, 1}) we find
\begin{equation}
\label{V, Model III}
\epsilon \, V = \frac{(n - 1)}{8}\, \left[2\,f' - n\,(a - f)^2\right].
\end{equation}
If we set $f = 0$ we regain the Einstein embeddings of Type II with $a = \pm \sqrt{C}$.

To complete the embedding we need to specify the function $f$. We wish to generalize the model given in \cite{Anderson}. To this end we choose 
\begin{equation}
\label{choive of f}
f(y) = - 2 \,p  \lambda \, (1 + \lambda\, y)^m,
\end{equation}
 where $(p, \, \lambda, \, m)$  are some constant parameters. In fact, substituting this into (\ref{Psi prime for example III}) we get

\begin{equation}
\label{Psi for Lambda = 0. Model I}
\Psi'^2 = - \frac{m\, p \,(n - 1)\,\lambda^2}{(1 + \lambda\,y)^{1 - m}}.
\end{equation}
When $m = - 1$ this expression reduces to (\ref{assumtion on F(y)}) and  $\Psi$ is given by (\ref{Psi for Lambda = 0}). For other values of $m$ the solution is

\begin{equation}
\label{Psi, Model I, m = neq - 1}
\Psi = \bar{q}\, (1 + \lambda \, y)^{(m + 1)/2} + \Psi_{0},\;\;\;\;m \neq - 1,
\end{equation}
where $\Psi_{0}$ is a constant of integration and 

\[
\bar{q} = \frac{2\, \sqrt{(- m\,p) (n - 1)}}{m + 1}. 
\]
The embedding function corresponding to (\ref{choive of f}) is given by
\begin{equation}
F(y) = F_{0}\, \mbox{e}^{a y}\, \times \exp \left[\frac{2 \,p \, (1 + \lambda\,y)^{m + 1}}{m + 1}\right], \;\;\;\; m \neq - 1,
\end{equation}
and 
\begin{equation}
F(y) = F_{0}\, \mbox{e}^{a y}\, \times (1 + \lambda\, y)^{2 p}, \;\;\;\; m  = -  1.
\end{equation}
For $m = - 1$ the scalar field is identical to (\ref{Psi for Lambda = 0}). However, the potential is not equal to (\ref{potential for Lambda = 0}) - unless $a = 0$.

For $a = 0$ and $m \neq 1$ the potential has the form 
\begin{equation}
\label{V for Model I and m neq - 1}
\epsilon \,V = - \frac{(n - 1)\, p\, \lambda^2}{2}\, [n\, p\, Z^{2 m} + m\, Z^{m - 1}], \;\;\;\;Z = \left(\frac{\Psi - \Psi_{0}}{\bar{q}}\right)^{2/(m + 1)}, \;\;\;\;m \neq - 1.
\end{equation}
The above examples demonstrate   that the embeddings provided by the metric (\ref{metric in (n + 1), after change of scale}) are consistent with higher-dimensional scalar fields with different potentials (not only exponential), in contrast with what is claimed in \cite{Anderson}.

\section{Final remarks}

We have shown that all vacuum solutions of general relativity in $n$-dimensions, with or without a  
cosmological constant $\Lambda_{(n)}$, can  in several nonequivalent ways  be embedded in a $(n + 1)$-dimensional space   which is either an Einstein space or a solution of the higher-dimensional field equations whose source is a  real massless scalar field $\Psi$, minimally-coupled to Einstein gravity and self-interacting through a potential $V(\Psi)$. This work generalizes a number of results discussed in the  literature \cite{MW1, MW2, MWL, Anderson, Fonseca, Bejancu}.

From (\ref{metric in (n + 1), after change of scale}) it follows that all the hypersurfaces orthogonal to the unit vector $\hat{n}^{A} = \delta^{A}_{y}$ inherit the line element

\begin{equation}
\label{the foliation inherit the line element}
\left.d S^2\right|_{\Sigma} = d s^2 = F(y)\, g_{a b} \, d x^a\, d x^b.
\end{equation}
 Since  ${^{(n)}R}_{\mu \nu}$ is invariant under a constant conformal transformation of the metric,  the effective cosmological constant in $n$-dimensions, say $\Lambda_{(n)}^{eff}$,  is 

\begin{equation}
\label{effective Lambda in n-dim}
\Lambda_{(n)}^{eff} = \frac{\Lambda_{(n)}}{F(y)}.
\end{equation} 
This is a consequence of the fact that the induced metric (\ref{the foliation inherit the line element}) has a scaling symmetry; if we  perform a coordinate transformation
\[
d \bar{x}^a = \sqrt{F}\, d x^a,
\]
then, in terms of the barred coordinates,  $d s^2 = {\bar{g}}_{a b}\, d \bar{x}^a\,  d \bar{x}^b$. The metric  
${\bar{g}}_{a b}= {\bar{g}}_{a b}(\bar{x})$ has the same form as $g_{a b}$ but with rescaled constants. As an illustration let us consider an example. For  simplicity we examine the Schwarzschild-de Sitter metric in spherical space
coordinates\footnote{This applies to more general solutions in vacuum, e.g., Kerr or Kerr-de Sitter in any number of dimensions \cite{Page}} ($r$, $\theta$, $\phi$)

\begin{equation}
\label{Schwarzschild-de Sitter metric in 4D}
d s^2 = h d t^2 - \frac{d  r^2}{h} - r^2\, (d \theta^2 + \sin^2 \theta d \phi^2),
\end{equation}
with
\begin{equation}
\label{h for the Schw-de Sitter}
h = 1 - \frac{2 M}{r} - \frac{\Lambda_{(4)}}{3}\,r^2,
\end{equation}
which is a solution of the four-dimensional Einstein field equations ${^{(4)}G}_{\mu \nu} = \Lambda_{(4)}\, g _{\mu \nu}$. The two parameters $M$ and $\Lambda_{(4)}$ represent  the mass of the central body and the cosmological constant, respectively.

The five-dimensional spacetime with metric

\begin{equation}
\label{extension to five dimensions of Schw-de Sitter}
d S^2 = \gamma_{A B}\, d x^{A} d x^{B} = F(y)\,\left(h d t^2 - \frac{d  r^2}{h} - r^2\, d \Omega^2\right) + \epsilon d y^2,
\end{equation}
where $F$ is any of the functions discussed in the previous sections, provides an embedding for the Schwarzschild-de Sitter metric. 
In fact, on any hypersurface $\Sigma$ orthogonal to $y$, after the transformation of coordinates 
\[
\bar{r} = \sqrt{F}\, r, \;\;\;\;\;\bar{t} = \sqrt{F}\, t
\]
the metric (\ref{extension to five dimensions of Schw-de Sitter})  reduces to 

\begin{equation}
\label{Schw-de Sitter with transformed M and Lambda}
\displaystyle \left. d S^2\right|_{\Sigma} = d s^2 = \left(1 - \frac{2 \, M^{eff}}{\bar{r}} - \frac{\Lambda_{(4)}^{eff}}{3}\, \bar{r}^2\right)\, d \bar{t}^2 - \left(1 - \frac{2 \, M^{eff}}{\bar{r}} - 
\frac{\Lambda_{(4)}^{eff}}{3}\, \bar{r}^2\right)^{- 1}\, d \bar{r}^2 - \bar{r}^2 \, d \Omega^2,
\end{equation}
where the effective mass $M^{eff}$ and cosmological constant ${\Lambda}^{eff}_{(4)}$ in 4D depend on the extra coordinate, viz.,
\begin{equation}
\label{effectime M and Lambda as functions of F}
M^{eff} = M \sqrt{F(y)}, \;\;\;\;\;\Lambda_{(n)}^{eff} = \frac{\Lambda_{(4)}}{F(y)}.
\end{equation}
If physics takes place on a hypersurface of constant $y$, then $M$  and $\Lambda_{(4)}$ are constants.  
However, if we adopt the approach used elsewhere, in which $y(s)$  is given by a solution of the 5D geodesic equation, then these quantities are no longer constants but vary with time \cite{MW1, MW2, JPdeLRussianJ, OWM}. This approach can bring to light important physical effects, e.g., in the early universe and galaxy formation. The study 
of such effects is beyond the scope of this work.

\medskip

Finally, we note that  a  nonvanishing logarithmic derivative of the embedding function $F$ gives rise to an extra (non-gravitational) force acting in $n$-dimensions. This follows from the assumption that massless particles in $(n + 1)$ move along null geodesics, in the same way as photons 
in 4D. Indeed, the geodesic equation in the embedding higher-dimensional space is

\begin{equation}
\label{geodesic equation in (n + 1)}
\frac{d^2 x^A}{d\chi^2} + K^{A}_{BC}\frac{dx^B}{d\chi}\frac{dx^C}{d\chi} = 0,
\end{equation}
where $\chi$ is some affine parameter along the geodesic. To obtain the  $n$-dimensional part of the geodesic equation (\ref{geodesic equation in (n + 1)}) we set $A = a = 0, \,1, ..., \, n$ and introduce the  function $l$ as 

\begin{equation}
\label{definition of l}
d\chi = l  d s.
\end{equation}
After some manipulations we get
\begin{equation}
\label{nD part of the geod. equation in terms of ds}
\frac{d^2 x^{a}}{d s^2} + \Gamma^{a}_{b c}\, \frac{d  x^{b}}{d s}\frac{d  x^{c}}{d s} =  \left[\frac{1}{l}\,\frac{d l}{d s} - \frac{F'}{F}\, \frac{d y}{d s}\right] \,\frac{d  x^{a}}{d s}. 
\end{equation}
Now setting $A = y$, and using (\ref{the foliation inherit the line element}),  we find
\begin{equation}
\label{equation for A = y}
\frac{d^2 y}{ds^2} - \frac{1}{l}\left(\frac{d l}{ds}\right)\left(\frac{dy}{ds}\right) - \frac{\epsilon \, F'}{2\, F} = 0.
\end{equation}
If $\epsilon = - 1$, then the assumption $d S^2 = 0$ implies $d s^2 = d y^2 > 0$ which   corresponds to a timelike geodesic of a massive particle in $n$-dimensions. Along such geodesics we can take $(d y/ d s) = 1$. Finally, combining (\ref{nD part of the geod. equation in terms of ds})-(\ref{equation for A = y}) we obtain

\begin{equation}
\label{fifth force}
\frac{d^2 x^{a}}{d s^2} + \Gamma^{a}_{b c}\, \frac{d  x^{b}}{d s}\frac{d  x^{c}}{d s} =  - \frac{F'}{2 F}\,\frac{d  x^{a}}{d s}. 
\end{equation}
The term on the right-hand side modifies the 
conventional geodesic motion. It represents an extra force per unit mass (a fifth force)   that causes an anomalous acceleration,  which in principle could  be detected experimentally.

\renewcommand{\theequation}{A-\arabic{equation}}
  \setcounter{equation}{0}  
  \section*{Appendix A: Evaluation of the Kretschmann scalar}  

To investigate the presence of singularities in the embedding spacetime we study the Kretschmann scalar 
\begin{equation}
\label{Kretschmann scalar in (n + 1)}
\widehat{K}^2 = {\widehat{R}}_{A B C D}\, {\widehat{R}}^{A B C D},
\end{equation}
where ${\widehat{R}}_{A B C D} = {^{(n + 1)}R}_{A B C D}$ is the Riemann tensor in $(n + 1)$ - dimensions. For the embeddings under consideration, generated by the line element (\ref{metric in (n + 1), after change of scale}), Christoffel symbols  (\ref{Christoffel symbols, 1}) reduce to 

\begin{eqnarray}
\label{Christoffel symbols, 2}
K_{y y}^{y} &=& K_{y y}^{a} = K_{a y}^{y} = 0\nonumber \\
K_{y a}^{b} &=& \frac{F'}{2 F}\, \delta_{a}^{b}, \;\;\;K_{a b}^{y} = - \frac{\epsilon}{2}\, F'\,g_{a b},\;\;\;K_{a b}^{c} = \Gamma_{a b}^{c}. 
\end{eqnarray}
As a consequence the Riemann tensor in $(n + 1)$-dimensions can be decomposed as

\begin{eqnarray} 
    {\widehat{R}}_{a b c d} &=& F \,R_{a b c d}\, + \epsilon\,\frac{F'^2}{4}\, \left(g_{b c} \,g_{a d} - g_{b d}\,  g_{a c}\right)\\
{\widehat{R}}_{a b c y} &=& 0 \\
{\widehat{R}}_{a y b y} &=& - \frac{1}{2}\, g_{a b}\,\left(F'' - \frac{F'^2}{2 F}\right).
\end{eqnarray}
Here $R_{a b c d} = {{^{(n)}}R}_{a b c d}$ is the Riemann curvature tensor constructed with the metric $g_{a b}$. To evaluate (\ref{Kretschmann scalar in (n + 1)}) we  note that  $\widehat{K}^2 = \left({\widehat{R}}_{a b c d}\, {\widehat{R}}^{a b c d} + 4 \, {\widehat{R}}_{a y b y}\, {\widehat{R}}^{a y b y}\right)$, with

\begin{eqnarray}
{\widehat{R}}_{a b c d}\, {\widehat{R}}^{a b c d} &=& \frac{1}{F^2}\, R_{a b c d}\, R^{a b c d} - \frac{\epsilon\,  F'^2}{ F^3}\, R + \frac{n (n - 1)}{8}\,\left(\frac{F'}{F}\right)^4\\
{\widehat{R}}_{a y b y}\, {\widehat{R}}^{a y b y} &=& \frac{n}{4 F^2}\, \left(F'' - \frac{F'^2}{2 F}\right)^2,
\end{eqnarray}
where $R$ is the scalar curvature of the $n$-dimensional space, i.e., $R = g^{a c}\, g^{b d}\, R_{a b c d}$. For the spacetime  described by  (\ref{Field equations in n-dim}) it is  $R = - [2\,n\,\Lambda_{(n)}/(n - 2)]$. Thus, using the above expressions we obtain 
\begin{equation}
\label{Kretschmann scalar in (n + 1), explicit expression}
\widehat{K}^2 = \frac{K^2}{F^2} + n \,\left[ 
\left(\frac{F''}{F}\right)^2 - \frac{F''\, F'^2}{F^3} + \frac{(n + 1)}{8}\,\left(\frac{F'}{F}\right)^4 + 
\frac{2\,  \epsilon\, \Lambda_{(n)}}{n - 2}\,\frac{F'^2}{F^3}\right],
\end{equation}
with $K^2 = R_{a b c d}\, R^{a b c d}$.

\end{document}